\documentstyle[twoside,fleqn,espcrc2]{article}

\input epsf

\newcommand{\AmS}{{\protect\the\textfont2
  A\kern-.1667em\lower.5ex\hbox{M}\kern-.125emS}}

\title{String tension and monopoles in $T \neq 0$ SU(2) QCD}

\author{Shinji Ejiri 
\address{
Department of Physics, Kanazawa University, Kanazawa 920-11, Japan}
\thanks{Presented by S. Ejiri}
, Shun-ichi Kitahara 
\addtocounter{address}{-1}
\addressmark
, Yoshimi Matsubara 
\address{
Nanao Junior College, Nanao, Ishikawa 926, Japan}
 and Tsuneo Suzuki
\addtocounter{address}{-2}
\addressmark
}

\begin{document}

\begin{abstract}
Monopole and photon contributions to abelian Wilson loops are calculated using 
Monte-Carlo simulations of finite-temperature $SU(2)$ QCD 
in the maximally abelian gauge. 
The string tension is reproduced by monopole contribution alone 
also in finite temperature SU(2) QCD.
The spatial string tension scales as 
$\sqrt{\sigma} \propto g^{2}(T)T$
and is reproduced almost by monopole contribution alone.
Each configuration has one long monopole loop,
and the long monopole loops alone are responsible for the string tension 
in the confinement phase.
On the other hand, the spatial string tension 
in the deconfinement phase is reproduced
by wrapped monopole loops alone.
\end{abstract}

\maketitle

\section{Introduction}

The dual Meissner effect due to 
condensation of color magnetic monopoles
is conjectured to be the color confinement mechanism in QCD.
We consider QCD after abelian projection\cite{thooft2}. 
The abelian projection of QCD is to extract an abelian theory 
performing  a partial gauge-fixing.

A gauge called maximally abelian (MA) gauge is 
interesting among many abelian projections\cite{kron}. 
The string tension can be reproduced 
from residual abelian link variables\cite{yotsu}.
Moreover, the string tension is explained 
by monopole contributions alone\cite{shiba2} 
as in compact QED \cite{stack}. 

The aim of this note is 
1)\ to show that the string tension is explained by monopole 
contribution alone also in finite-temperature $SU(2)$ QCD, 
2)\ to study the spatial string tension 
both in the confinement and in the deconfinement phases and
3)\ to study what kind of monopole loops are responsible for 
the physical and the spatial string tensions.

\section{Formalism and simulations}

We adopt the usual $SU(2)$ Wilson action 
and choose the maximally abelian  gauge 
in which diagonal components of all link variables are maximized.
Gauge fixed link variables are decomposed into a product of two matrices:
\mbox{$ \widetilde{U}(s,\hat\mu) = c(s,\mu)u(s,\mu) $}
where $u(s,\mu)$ is diagonal abelian gauge field. 
An abelian Wilson loop operator is 
given  by a product of monopole and photon contributions\cite{shiba2}. 
\begin{eqnarray}
W\    & \hspace{-3.2mm} = \hspace{-3.2mm} & W_{1} \cdot W_{2} \label{w12}\\
W_{1} & \hspace{-3.2mm} = \hspace{-3.2mm} & 
\exp\{-i \sum \partial'_{\mu}\bar{f}_{\mu\nu}(s)
D(s-s')J_{\nu}(s')\} \nonumber \\
W_{2} & \hspace{-3.2mm} = \hspace{-3.2mm} & \exp\{2\pi i \! \sum 
\! k_{\beta}(s)D(s \! - \!s' \!)\frac{1}{2}
\epsilon_{\alpha\beta\rho\sigma}\partial_{\alpha}M_{\rho\sigma}( \! s'\! )\}, 
\nonumber
\end{eqnarray}
where $\partial$ $(\partial')$ is a forward (backward) 
derivative on the lattice, 
$J_{\mu}(s)$ is an external current taking $\pm 1$ along the Wilson loop 
and $M_{\mu\nu}(s)$ is an antisymmetric variable 
as $J_{\nu}(s)=\partial_{\mu}'M_{\mu\nu}(s)$ also
\mbox{$f_{\mu\nu}(s)=\bar{f}_{\mu\nu}(s)+2\pi n_{\mu \nu}
=\partial_{\mu}\theta_{\nu}(s)-\partial_{\nu}\theta_{\mu}(s)$} 
$(-\pi <\bar{f}_{\mu \nu}(s) \leq \pi)$, where 
$\theta_{\mu}(s)$ is the angle variable defined from  $u(s,\hat\mu)$. 
$D(s)$ is the lattice Coulomb propagator and 
a monopole current $k_{\mu}(s)$ is defined as $k_{\mu}(s)= 
(1/4\pi)\epsilon_{\mu\alpha\beta\gamma}\partial_{\alpha}
\bar{f}_{\beta\gamma}(s)$ following DeGrand-Toussaint\cite{degrand}.
$W_{1} (W_{2})$ 
is the photon (the monopole) contribution to the abelian Wilson loop.
To study the features of both contributions, 
we calculated the expectation values of $W$ (called abelian), 
of $W_1$ (photon part) and of $W_2$ (monopole part), separately. 

The Monte-Carlo simulations were performed on $24^3\times 8$ lattice 
at $\beta =2.3 \sim 2.8$. 
All measurements were done every 50 sweeps after
a thermalization of 2000 sweeps. We took 50 
configurations totally for measurements. 
Assuming the static potential is given 
by linear + Coulomb + constant terms, 
we tried to determine the potential by the Creutz ratios 
using the least square fit.

\begin{figure}[tb]
\vspace{-10mm}
\epsfxsize=0.4\textwidth
\begin{center}
\leavevmode
\epsfbox{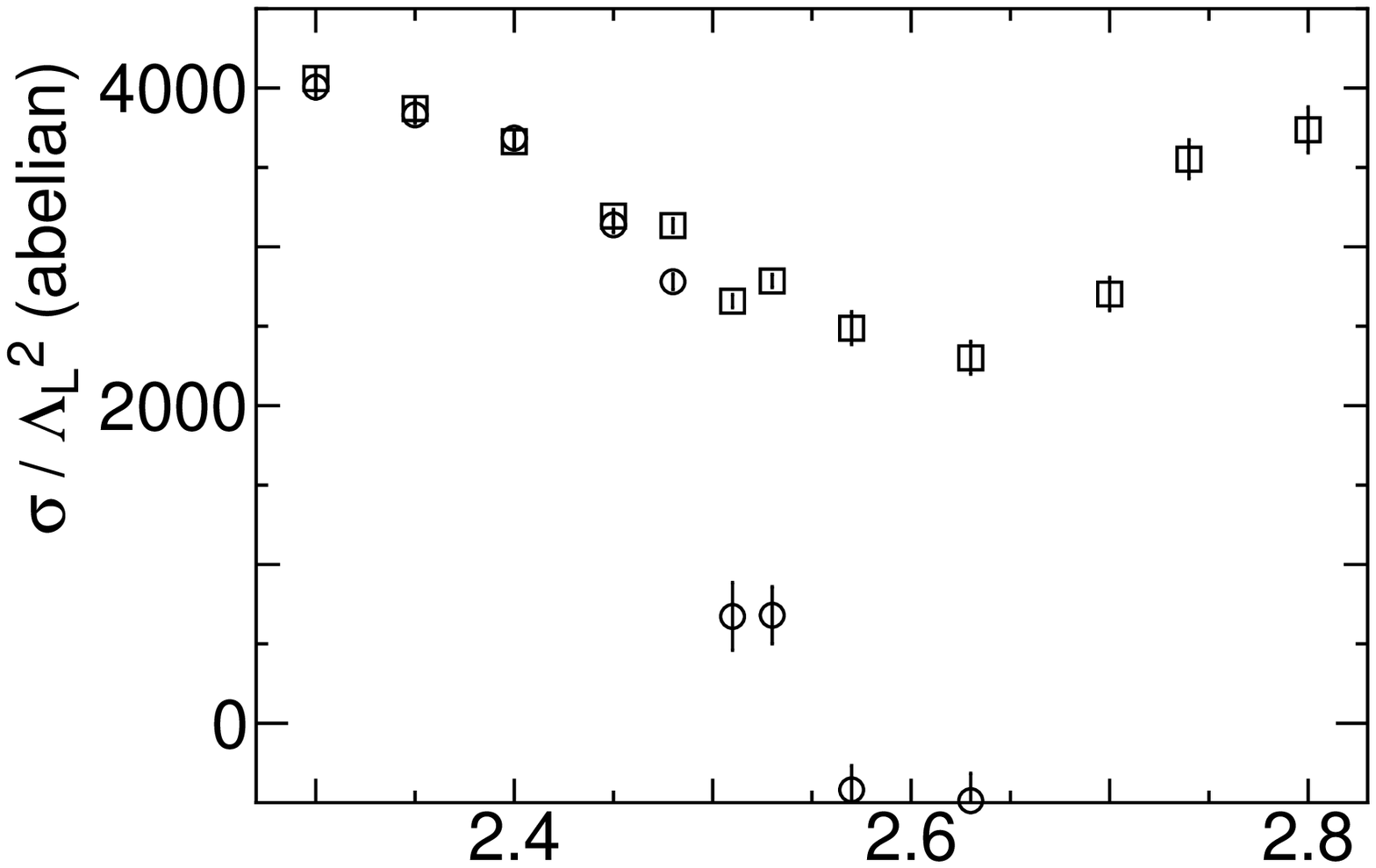}
\end{center}
\vspace{-37mm}
\epsfxsize=0.4\textwidth
\begin{center}
\leavevmode
\epsfbox{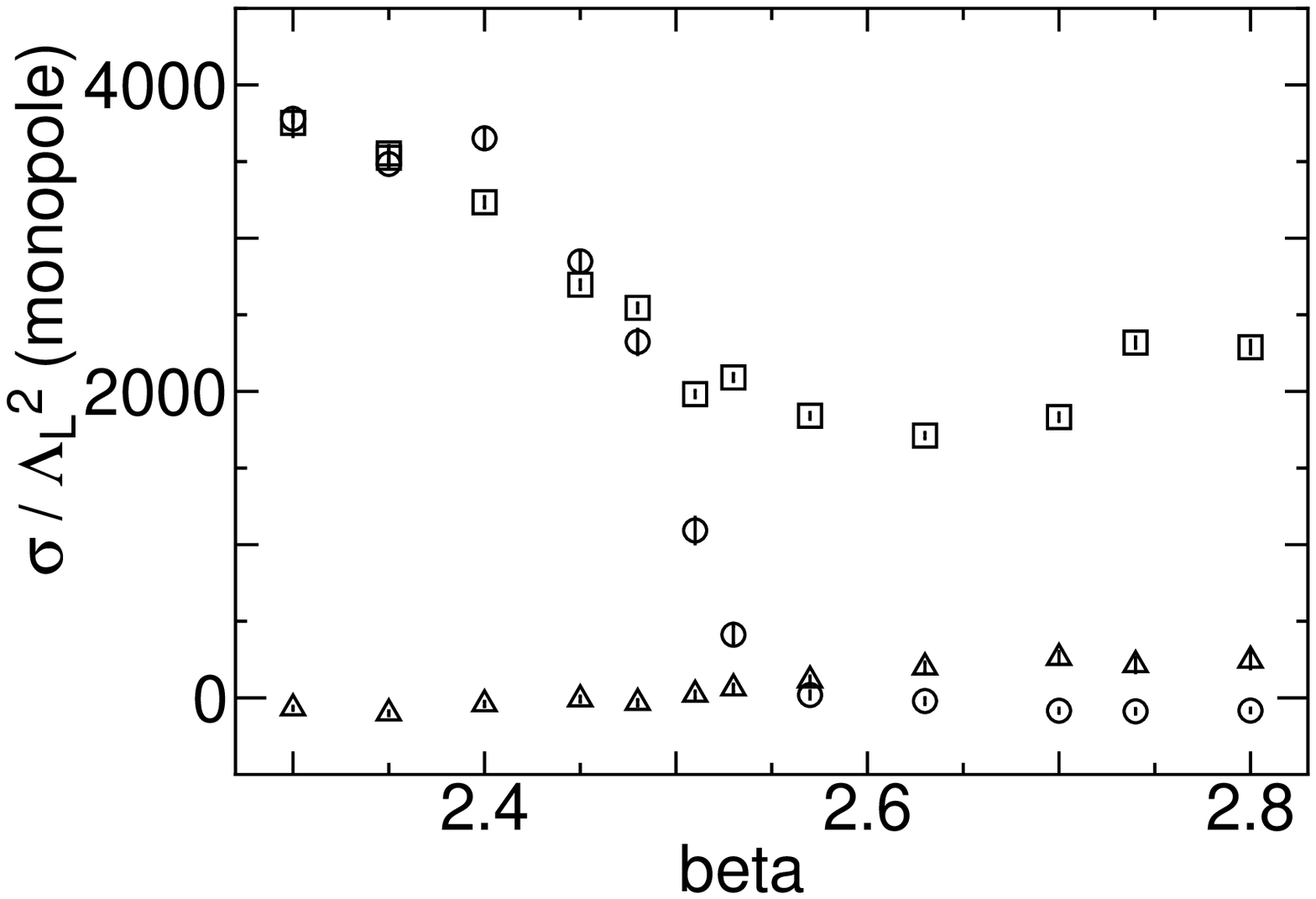}
\end{center}
\vspace{-32mm}
\caption{
Physical string tensions (circle) and spatial string tensions (square)
from the abelian Wilson loops (upper) and from the monopoles (lower).
The triangles are physical string tensions from the photons.}
\label{st8}
\vspace{-5mm}
\end{figure}

The data are shown in Fig.\ \ref{st8}.
We calculated the physical string tension and 
the spatial string tension. 
The latter is given by Wilson loops composed of 
only spatial link variables.
In the confinement phase, both string tensions 
from the abelian Wilson loops show the same value as that 
in ($T=0$) $SU(2)$ QCD,
which is also the same as the full string tension\cite{shiba2}.
Moreover, the physical string tension vanishes at the critical 
coupling $\beta_{c}$.
However, the spatial string tension does not vanish and remains finite even 
in the deconfinement phase. 
Both physical and spatial string tensions 
from the monopoles almost agree with those from the abelian Wilson loops
in the confinement phase.
In the deconfinement phase, the monopole contributions to the 
physical string tension vanish, whereas those to the spatial one remain 
non-vanishing.
The string tension from the photons is negligibly small.

\section{Scaling properties of the spatial string tension for $T > T_c$}

It is believed that, at high temperature, four dimensional 
QCD can be regarded through dimensional reduction as an effective 
three dimensional QCD with $A^0$ as a Higgs field\cite{pisarski}. 
In this effective theory,
the spatial string tension is expected to obey 
$ \sqrt{\sigma_s (T)} \propto g^2(T)T, $
where $g(T)$ is the four dimensional coupling constant.
The scaling properties of the spatial string tension 
derived from the usual full Wilson loops 
is confirmed recently in Monte-Carlo simulations of 
$SU(2)$ QCD\cite{bali}.   

To check if the same thing happens in the abelian case,
we performed additional Monte-Carlo simulations varying both 
$N_t$ and $\beta$ on $24^{3} \times N_t$ lattices
following Bali et al.\cite{bali}.
We measured the string tension for $N_t = 2, 4, 6, 8 \ {\rm and}
\ 12$ at $\beta= 2.30, 2.51 \ {\rm and} \ 2.74$ 
which are the critical points for $N_t = 
4, 8 \ {\rm and} \ 16$, respectively. 

\begin{figure}[tb]
\vspace{-10mm}
\epsfxsize=0.4\textwidth
\begin{center}
\leavevmode
\epsfbox{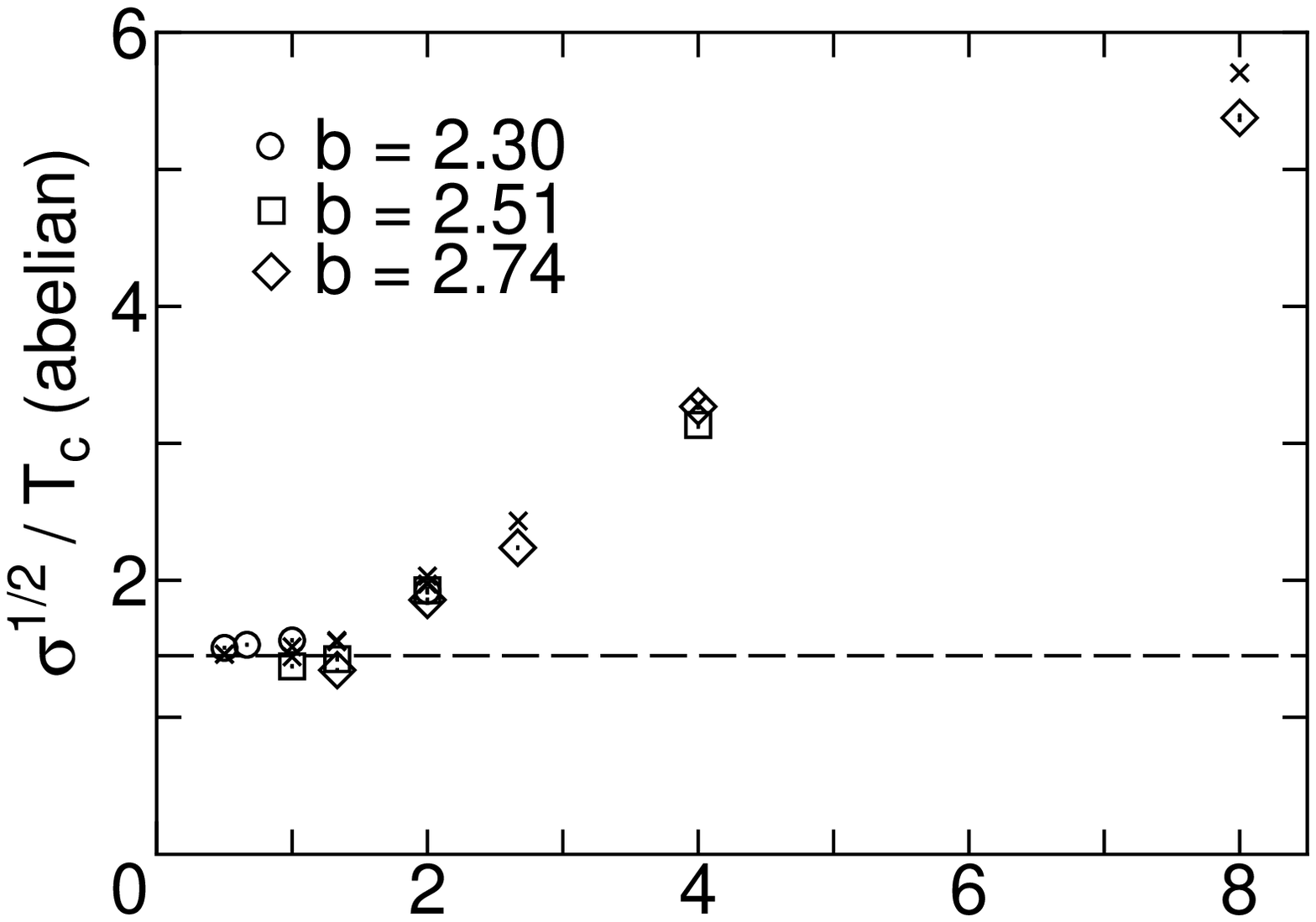}
\end{center}
\vspace{-37mm}
\epsfxsize=0.4\textwidth
\begin{center}
\leavevmode
\epsfbox{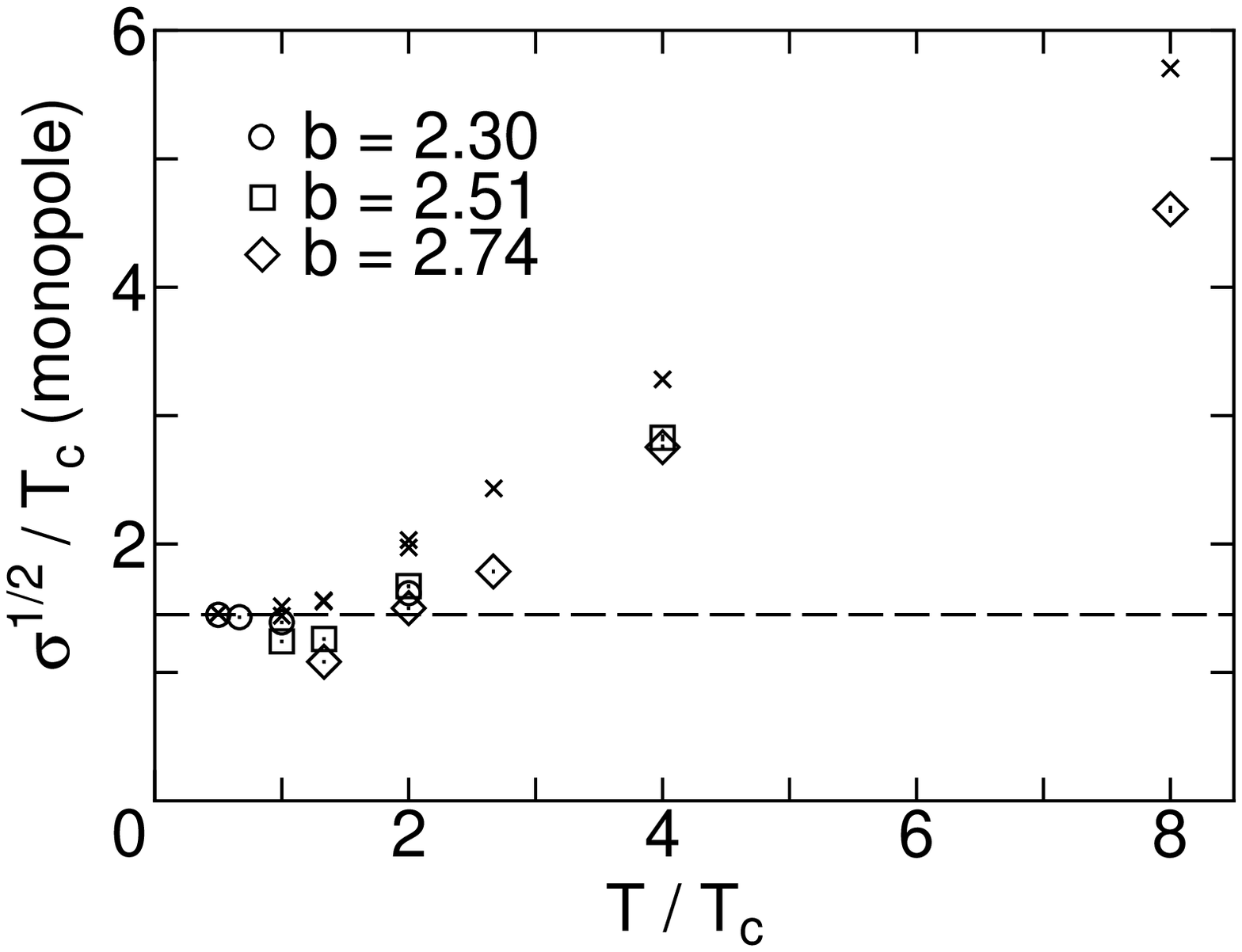}
\end{center}
\vspace{-32mm}
\caption{The spatial string tensions from the abelian Wilson loops (upper) 
and from the monopoles (lower) versus temperature. 
The crosses denote full spatial string tensions.
}
\label{spst}
\vspace{-5mm}
\end{figure}

The data are plotted in Fig.\ \ref{spst}. 
In the case of the spatial string tension derived 
from the abelian Wilson loops, 
we get almost the same behaviors as those of the full ones 
denoted by cross points. The latter is cited from Ref.\cite{bali}.
$\sqrt{ \sigma } / T_{c}$ is independent of $\beta$ and so the spatial string
tension is expected to be a physical quantity 
remaining in the continuum limit.
The spatial string tension from the monopoles is also $\beta$ independent, 
and is a little bit lower, but shows almost the same behavior.

\section{A long monopole loop, a wrapped monopole loop 
and the string tension}

 The features of the abelian monopole currents were studied in \cite{kita}
through Monte-Carlo simulations. They obtained the followings:
1) In the confinement phase, there is one long monopole loop 
in each configuration.
2) In the deconfinement phase, long monopole loops disappear. 
All monopole loops are short.

\begin{figure}[tb]
\vspace{-10mm}
\epsfxsize=0.4\textwidth
\begin{center}
\leavevmode
\epsfbox{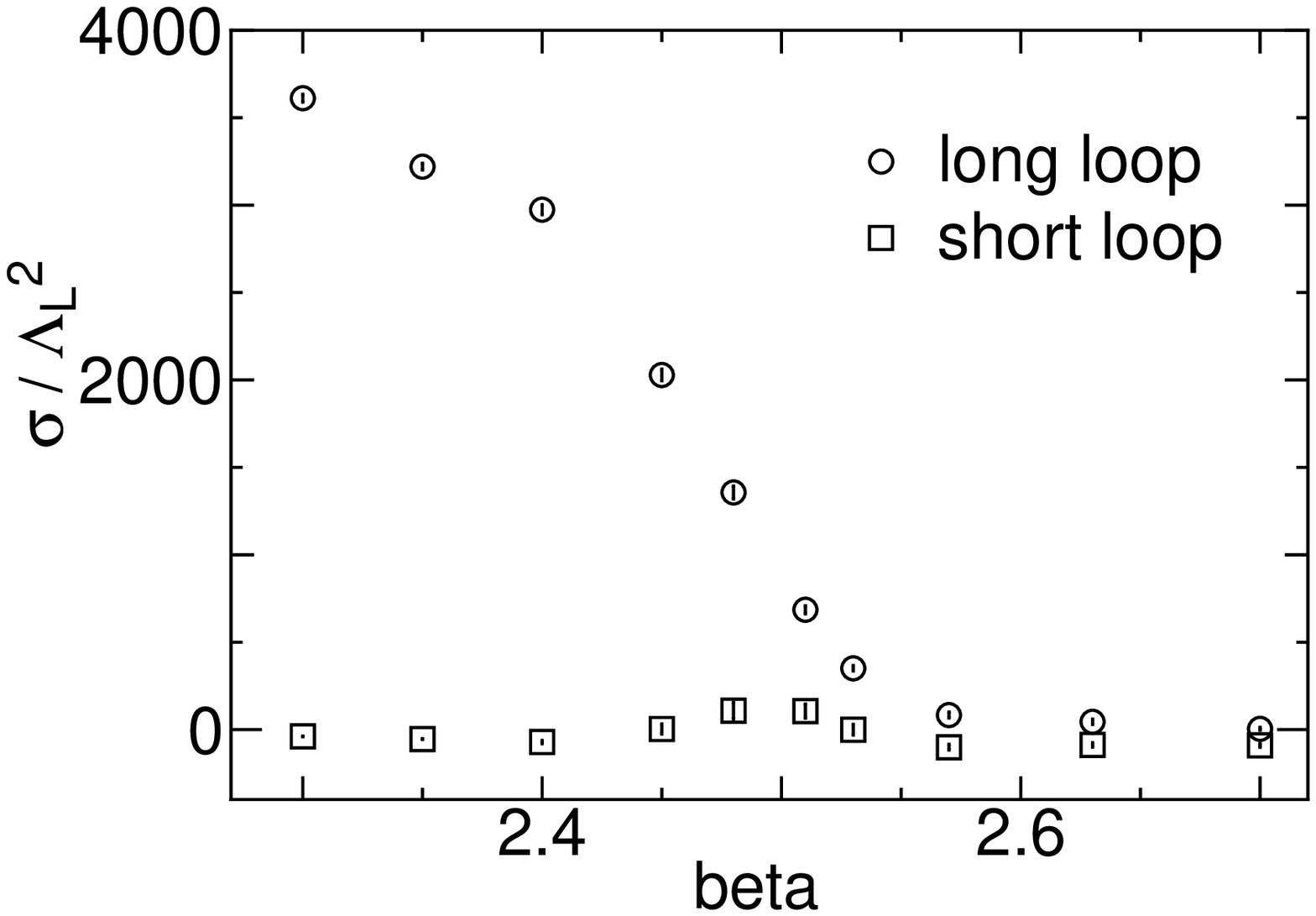}
\end{center}
\vspace{-33mm}
\caption{
The contributions from the longest monopole loop (circle) and 
 from all other monopole loops (square) to the string tensions.
}
\label{lloop}
\vspace{-19mm}
\end{figure}

These results bring us an idea that 
the long monopole current plays an important role in the string tension,
because the physical string tension exists only in the confinement phase and 
vanishes in the deconfinement phase.
We investigated contributions from the longest monopole loops and 
from all other monopole loops  to the string tension separately.
The data on $24^{3} \times 8$ lattice are shown in Fig.\ \ref{lloop}. Clearly, 
the contributions from the long loop alone  reproduce almost the full
value of the string tension. On the other hand, the short loop contributions 
are almost zero. 
Only a few percent of the total links
are occupied by monopole currents belonging to the long loop.
Nevertheless, it gives rise to the full value of the string tension.

\begin{figure}
\epsfxsize=0.4\textwidth
\begin{center}
\leavevmode
\epsfbox{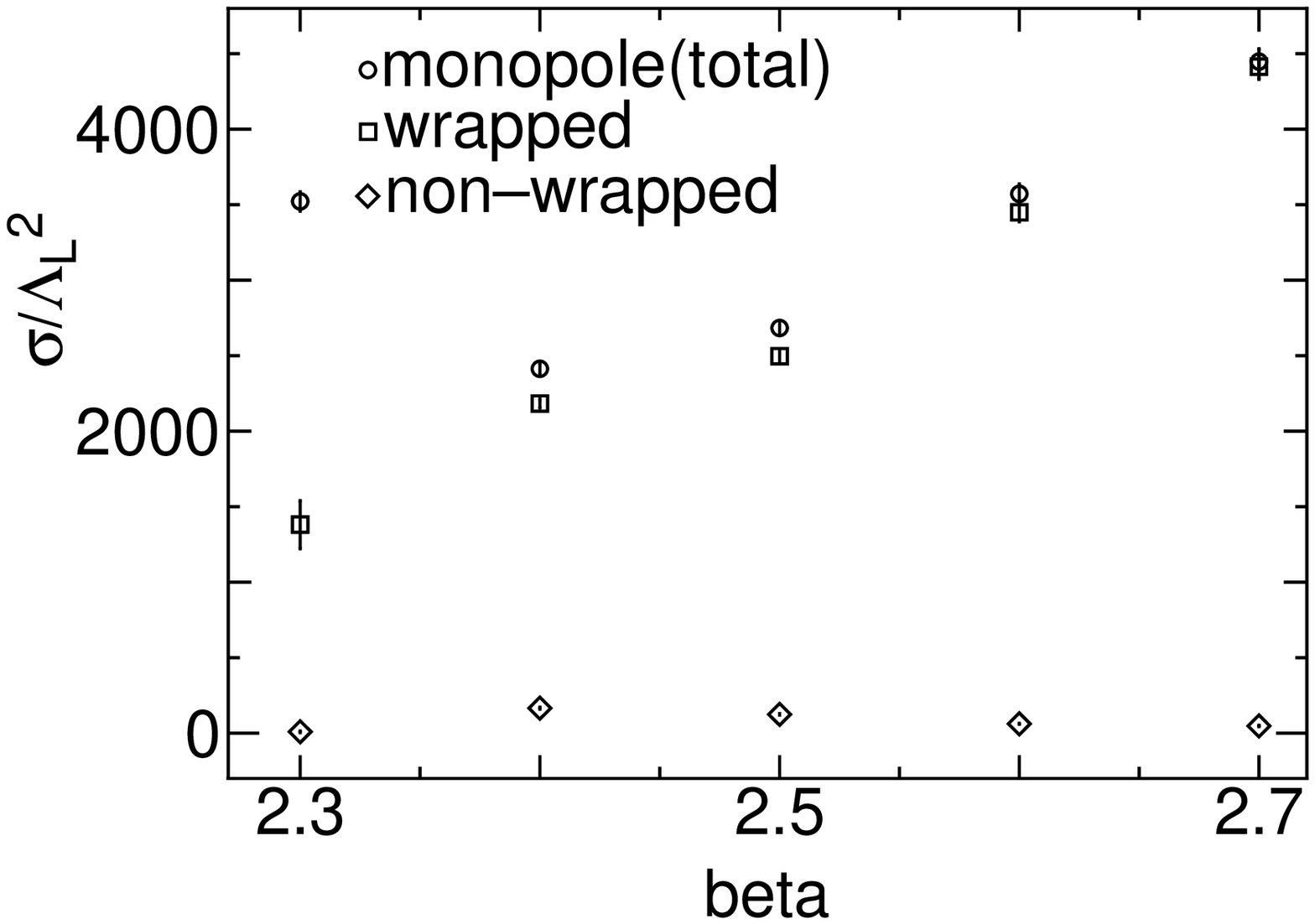}
\end{center}
\vspace{-32mm}
\caption{
The contributions from the wrapped loops (square), 
from the non-wrapped loops (diamond) and from the total monopole loops 
(circle) to the spatial string tensions.}
\label{wrap}
\vspace{-5mm}
\end{figure}

Next, we consider the case of the spatial string tension at high temperature.
The static (wrapped) monopole 
, which closes by periodicity of the time direction, 
is expected to be important for the spatial string tension, 
because, in three dimensional SU(2) QCD with a Higgs field, 
the confinement mechanism is explained by an instanton \cite{poly}, 
and this instanton is a static monopole 
in terms of four dimensional QCD at high temperature.
We calculated the contributions from the wrapped loops 
and the non-wrapped loops to the string tension separately.
The data on $16^{3} \times 4$ lattice are plotted in the Fig. \ref{wrap}.
About $ 30 \% $ of the monopole currents are the wrapped monopole currents
in deconfinement phase. 
The contributions from the wrapped loop alone reproduce almost the full
value of the spatial string tension from the total monopoles,
whereas those from the non-wrapped loops are almost zero.

Details are published in \cite{eji}.
This work is financially supported by JSPS Grant-in Aid for Scientific 
Research (B)(No.06452028).

\begin{thebibliography}{9}

\bibitem{thooft2} G. 'tHooft, Nucl. Phys. {\bf B190}, (1981) 455.
\bibitem{kron} A.S. Kronfeld et al., Phys. Lett. {\bf B198}, 
(1987) 516;
A.S. Kronfeld et al., Nucl.Phys. {\bf B293}, (1987) 461.
\bibitem{yotsu} T. Suzuki and I. Yotsuyanagi, 
Phys. Rev. {\bf D42}, (1990) 4257.
\bibitem{shiba2} H.Shiba and T.Suzuki, 
Phys. Lett. {\bf B333}, (1994) 461.
\bibitem{stack} J.D. Stack and R.J. Wensley, 
Nucl. Phys. {\bf B371}, (1992) 597.
\bibitem{degrand} T.A. DeGrand and D. Toussaint, 
Phys. Rev. {\bf D22}, (1980) 2478.
\bibitem{kita} S.Kitahara et al., in preparation.
\bibitem{pisarski} T.Appelquist and R.D.Pisarski, 
Phys. Rev. {\bf D23}, (1981) 2305.
\bibitem{bali} G.S.Bali et al., Phys. Rev. Lett. {\bf 71}, (1993) 3059.
\bibitem{poly} Polyakov, Nucl. Phys. {\bf B120}, (1977) 429.
\bibitem{eji} Ejiri et al., to appear in Phys. Lett. B.

\end{thebibliography}
\end{document}